\documentclass[sigconf]{acmart}

\usepackage{booktabs} 
\usepackage{todonotes}
\usepackage{microtype}
\usepackage{xspace}
\usepackage{balance}
\usepackage{paralist}
\usepackage{algorithm}
\usepackage{algpseudocode}
\usepackage{caption}
\usepackage{subcaption}
\usepackage[compact]{titlesec}

\renewcommand{\paragraph}[1]{\vspace{0.08em}\noindent {\bf #1}}

\AtBeginDocument{%
  \providecommand\BibTeX{{%
    \normalfont B\kern-0.5em{\scshape i\kern-0.25em b}\kern-0.8em\TeX}}}

\copyrightyear{2021}
\acmYear{2021}
\setcopyright{acmcopyright}\acmConference[ITiCSE 2021]{26th ACM Conference on Innovation and Technology in Computer Science Education V. 1}{June 26-July 1, 2021}{Virtual Event, Germany}
\acmBooktitle{26th ACM Conference on Innovation and Technology in Computer Science Education V. 1 (ITiCSE 2021), June 26-July 1, 2021, Virtual Event, Germany}
\acmPrice{15.00}
\acmDOI{10.1145/3430665.3456344}
\acmISBN{978-1-4503-8214-4/21/06}

\newcommand{\passedOriginalModel}{{.17}\xspace}
\newcommand{\passedOriginalAll}{{.024}\xspace}
\newcommand{\passedOriginalSeventy}{{.013}\xspace}
\newcommand{\passedOriginalNinety}{{.013}\xspace}

\newcommand{\passedSeventyNinety}{{.32}\xspace}

\newcommand{\emptyOriginalNinety}{{<.001}\xspace}

\newcommand{\emptySeventyNinety}{{.030}\xspace}

\newcommand{\deadOriginalNinety}{{<.001}\xspace}

\newcommand{\ApassedOriginalModel}{{.59}\xspace}
\newcommand{\ApassedOriginalAll}{{.66}\xspace}
\newcommand{\ApassedOriginalSeventy}{{.33}\xspace}
\newcommand{\ApassedOriginalNinety}{{.33}\xspace}

\newcommand{\ApassedSeventyNinety}{{.43}\xspace}

\newcommand{\AemptySeventyNinety}{{.36}\xspace}

\newcommand{\meanPassedOriginal}{{12.36}\xspace}

\newcommand{\meanPassedNinety}{{16.64}\xspace}

\newcommand{\meanEmptyOriginal}{{.22}\xspace}

\newcommand{\meanEmptyNinety}{{1.25}\xspace}
\newcommand{\meanDeadOriginal}{{2.19}\xspace}

\newcommand{\meanDeadNinety}{{.03}\xspace}
\newcommand{\corrBlocksHints}{{.45}\xspace}
\newcommand{\corrPassedHints}{{.05}\xspace}
\newcommand{\PcorrBlocksHints}{{.006}\xspace}
\newcommand{\PcorrPassedHints}{{.773}\xspace}

\begin{document}
\fancyhead{}

\title{Guiding Next-Step Hint Generation Using Automated Tests}


\settopmatter{authorsperrow=3}
 
\author{\mbox{Florian Obermüller}}
\email{obermuel@fim.uni-passau.de}
\affiliation{%
  \institution{University of Passau}
  \city{Passau}
    \country{Germany}
}

\author{Ute Heuer}
\email{ute.heuer@uni-passau.de}
\affiliation{%
	\institution{University of Passau}
	\city{Passau}%
	\country{Germany}
}

\author{Gordon Fraser}
\email{gordon.fraser@uni-passau.de}
\affiliation{%
  \institution{University of Passau}
  \city{Passau}
  \country{Germany}
}
\renewcommand{\shortauthors}{Obermüller, et al.}

\begin{abstract}
%
%
Learning basic programming with \scratch can be hard for novices and tutors
alike: Students may not know how to advance when solving a task, teachers may
face classrooms with many raised hands at a time, and the problem is
exacerbated when novices are on their own in online or virtual lessons.
%
%
It is therefore desirable to generate next-step hints automatically to provide
individual feedback for students who are stuck, but current approaches rely on
the availability of multiple hand-crafted or hand-selected sample solutions
from which to draw valid hints, and have not been adapted for \scratch.
%
%
Automated testing provides an opportunity to automatically select suitable
candidate solutions for hint generation, even from a pool of student solutions
using different solution approaches and varying in quality.
%
%
In this paper we present \textsc{Catnip}, the first next-step hint generation
approach for \scratch, which extends existing data-driven hint generation
approaches with automated testing.
%
%
Evaluation of \textsc{Catnip} on a dataset of student \scratch programs
demonstrates that the generated hints point towards functional improvements, and the use of automated tests allows the hints to be better individualized for the chosen solution path.
\end{abstract}



\newcommand{\litterbox}{\textsc{LitterBox}\xspace}
\newcommand{\scratch}{\textsc{Scratch}\xspace}
\newcommand{\drscratch}{\textsc{Dr. Scratch}\xspace}
\newcommand{\hairball}{\textsc{Hairball}\xspace}
\newcommand{\qualityhound}{\textsc{Quality Hound}\xspace}
\newcommand{\findbugs}{\textsc{FindBugs}\xspace}
\newcommand{\catnip}{\textsc{Catnip}\xspace}
\newcommand{\whisker}{\textsc{Whisker}\xspace}
\newcommand{\bastet}{\textsc{Bastet}\xspace}
\newcommand{\itch}{\textsc{Itch}\xspace}
\newcommand{\isnap}{\textsc{iSnap}\xspace}
\newcommand{\snap}{\textsc{Snap!}\xspace}
\newcommand{\sourcecheck}{\textsc{SourceCheck}\xspace}
\newcommand{\hintfactory}{\textsc{Hintfactory}\xspace}


\maketitle

\section{Introduction}
\label{sec:introduction}

Learning to program can be challenging and frustrating \cite{hansen2007}, therefore block-based programming
languages like \scratch~\cite{maloney2010} aim to remove some common obstacles,
such as the need to memorize programming commands and to produce complex but
syntactically valid textual structures~\cite{mcgill2020}. Nevertheless
acquiring confidence and skills in programming requires plenty of exercise and
time. In a classical learning environment novices have a tutor they may ask for
help if they are stuck solving an exercise. However, in large classes this task
can be challenging for the tutor if multiple children have questions at the
same time. Furthermore the emergence of, and recent need for, online courses
and virtual, asynchronous learning leads to a need for alternatives.
Consequently, it is desirable to automatically create feedback and individualized hints.

A common approach for generating next-step hints is to use data on possible
solutions for the task at hand, trying to identify the best candidate solution
matching the current location the student is working on, and then generating a
suggestion that moves the student solution closer to the chosen candidate
solution. Thus, these data-driven hint generation approaches have two
requirements: First, a suitable set of candidate solutions selected or created
by a tutor, and second an implementation of a differencing algorithm that can
extract concrete edit suggestions for a problem at hand using these candidate
solutions. The former is usually solved with manual labour. There are tools
solving the latter challenge, but to the best of our knowledge there are none
yet for \scratch.

In this paper we introduce \catnip, a new individual hint generation tool for
\scratch. \catnip adapts ideas for data-driven
hint-generation~\cite{zimmerman2015,price2017a} for usage in the \scratch
context. In order to increase automation further and to allow better
individualization of hints, \catnip uses automated
tests~\cite{stahlbauer2019testing} to select candidate solutions from
pools of student programs. An automated test sends events to a program and
observes its reactions, thereby determining whether it implements desired
behavior. The more such tests a program passes, the more likely it is suited
as a candidate solution for hint suggestions.

Evaluation on a set of student solutions to a non-trivial \scratch game
demonstrates that \catnip is able to provide high-quality suggestions to
student programs which lead to an improvement in functionality and code
quality. Our evaluation further demonstrates the importance of functional
correctness and diversity of candidate programs: If sufficient functionality
and diversity of the candidate programs are not enforced, then the generated
hints may be counterproductive. 
Consequently, \catnip's use of automated tests represents a fundamental aspect
of an automated, high quality solution for next-step hint generation.


\section{Background}\label{sec:background}

\scratch\cite{maloney2010} is a widely used visual block-based programming language that aims to make programming more accessible for novices. \scratch emphasizes exploration over recall and visually distinguishes different categories of blocks. It also features intuitive interactivity via an event driven paradigm  and offers simple media integration~\cite{bau2017}. Blocks are divided into different shapes, stackable blocks (also called statements) and reporter blocks that can fit inside stackable blocks (also called expressions). The blocks prevent syntactical errors so novices can focus very intuitively on what the program should do, but it can nevertheless be challenging to produce combinations of blocks that achieve the intended behavior.

Unsuitable challenges can frustrate learners, leading to a lack of interest and consequently to disengagement concerning learning tasks~\cite{renninger2010working}. To level challenges so that students are more likely to persevere, hints as part of a scaffolding process can be used~\cite{reiser_tabak_2014}. Producing hints automatically helps students and teachers alike. The former receive instant advice, and no face-to-face help-seeking is needed, thus avoiding peer stigmatisation~\cite{collins_kapur_2014}. The latter are able to mediate classroom management more concisely.    

There are different approaches to automatically produce various types of hints:

\subsection{Hints on Code Smells in \scratch}\label{sec:2.1.1}

Code smells are idioms that increase the likelihood of bugs in
code~\cite{fowler1999} during future modifications. Code smells for \scratch
have been derived from other programming languages \cite{hermans2016a} or
specially constructed for block-based languages~\cite{moreno2014}. It has been
established that code smells have negative effects on
learning~\cite{hermans2016b} and the \scratch
community~\cite{techapalokul2017a}. In order to derive information about code
smells in \scratch projects there are automated tools like
\hairball\cite{boe2013}, \qualityhound\cite{techapaloku2017b}, or
\litterbox~\cite{fraedrich2020}. While hints on code smells can help to improve code quality and prevent future bugs, they do not address the problem of learners being stuck.

\subsection{Hints on Bug Patterns in \scratch}\label{sec:2.1.2}

A bug pattern is a code idiom that is likely to be a
defect~\cite{hovemeyer2004}. The \litterbox tool~\cite{fraedrich2020} can
detect bug patterns (in addition to code smells) in \scratch programs. It tags
their occurrence in the \scratch code and displays generic hints on why there is a bug pattern and how to fix it. 
If learners are stuck because their programs contain bugs,
then hints on bug patterns provide actionable information that can help to
remove the bugs and proceed with the overall programming task. However, if
learners have arranged blocks in an incorrect way that does not match any
known bug pattern, or they are simply stuck because they do not know what to
do next, then bug patterns cannot serve as appropriate hints.

\subsection{Hints on Correctness in \scratch}\label{sec:2.1.3}

The functionality of a \scratch program can be checked against the desired
functionality using automated testing tools like \itch~\cite{johnson2016itch}
and \whisker~\cite{stahlbauer2019testing}, or model checking tools like
\bastet~\cite{stahlbauer2020}.
To utilize these approaches, educators need to either create test cases
manually, i.e., they need to describe the order in which input should be
sent to the program under test and what the expected response of the program
should be, or they have to formalize a specification of the expected behavior.
A common application of these approaches is automatic
grading~\cite{stahlbauer2019testing}. In general, a fundamental challenge is
that writing the necessary specifications even for school exercises can be a
tedious task; furthermore, while these approaches can point out problems in
programs, they do not provide suggestions on how to overcome these problems.


\subsection{Hints on Next Steps in Programming}

A possible approach to point out not only deficiencies in the code but also
their remedies lies in explicitly modelling the possible errors that can be
made in a given task. For example, Singh et al. \cite{singh2012} created such a
system utilizing a sample solution and an error model written by the tutor to
generate feedback for Python programming exercises.

In order to avoid the effort of having to explicitly model the possible errors,
Zimmerman and Rupakheti~\cite{zimmerman2015} proposed to use multiple sample solutions
to determine the solution that is closest to the one a novice is seeking help
for, and then to use this for hint generation. 
The ``closeness'' between programs is measured using pq-gram distance~\cite{augusten2005} as an approximation of tree edit distances on the abstract syntax trees (ASTs) of the programs.

Instead of sample solutions, the Contextual Tree Decomposition (CTD) algorithm~\cite{price2016} uses log data of students  trying to solve a specific task. It generates an interaction graph that represents students actions, program states and transitions between them. Probabilities are annotated and may guide individual hint production for every node in the AST of a student solution until an acceptable program state is reached.
The CTD algorithm was implemented in \isnap\cite{price2017a}, a next step hint
generation tool for the \snap~\cite{garcia2015} block based programming
language. Since acquiring the required log data is challenging, the algorithm
used in \isnap has since changed to \sourcecheck~\cite{price2017b}, which is
similar to the approach of Zimmerman and Rupakheti~\cite{zimmerman2015}. First
the student's current code is matched to solutions, but in this case previous
correct student solutions are used instead of samples created by a tutor.
Second, the edits that bring the current snapshot closer to a solution are
calculated by mapping the two ASTs and identifying deletions, insertions, moves
and reorders.

Despite this prior work on next-step hint generation, to the best of our knowledge no next-step hint generation approaches have been implemented for \scratch. While in principle all of the above approaches could be adapted to \scratch, they all have challenging prerequisites:
Preparing error models or sample solutions is challenging since multiple different task solutions are required in order to generate sensible feedback. Without multiple solutions, hints would lead every student to a similar solution and thus ignore students differing in preferences, performance, skills and acquired concepts, contradicting the principle of individual feedback \cite{robins_2019}.
The CTD algorithm requires log data which is not easy to acquire, would require adaptation of the \scratch virtual machine, and would also require that students use such a modified version of \scratch rather than the official \scratch website.
The \sourcecheck~\cite{price2017b} algorithm tries to deal with this problem by
using student solutions from prior courses, but the correctness and
suitability of these solutions has to be checked by hand to guarantee that
the generated hints really lead to the desired behavior.



\section{\catnip: Next-Step Hints for \scratch}\label{sec:catnip}

\catnip is a next-step hint generation tool for \scratch. It takes one source
project (i.e., the current program of the student using the tool) and generates
hints to guide the project to become closer to a selected target project. It is
built on Zimmerman's and Rupakheti's hint generation
algorithm~\cite{zimmerman2015}, but makes several adaptations for the
application to \scratch. Furthermore, \catnip does not require the user to
provide hand-tailored sample solutions or error models, but instead selects
suitable candidate solutions from a pool of solutions using an automated
\whisker~\cite{stahlbauer2019testing} test suite. The overall process of hint
generation consists of four steps, which will be explained in more detail in
the remainder of this section:
\begin{enumerate}
	\item Identifying suitable candidate solutions for hint extraction
	\item Selecting the best matching candidate solution as target solution
	\item Determining AST differences
	\item Synthesizing hints
\end{enumerate}

\subsection{Identifying Suitable Candidate Solutions}
\label{subsec:TargetCandidates}

A prerequisite for data-driven hints is a selection of suitable candidate projects from which to extract possible edits. Traditionally, hints are generated using candidate solutions manually created or selected by an instructor. To reduce the effort of this selection, \catnip can automatically select suitable candidates from a pool of prior solutions to the same task. Since incorrect solutions would lead to counterproductive hints, \catnip relies on automated tests in order to decide which solutions from the pool to use as candidates for hint generation. The user has to specify the pool of solutions, a \whisker test suite, as well as a threshold on the percentage of tests that need to pass in order for a solution to qualify as candidate for hint generation; the default value for this threshold is 90\%, such that only very good solutions are included.

\subsection{Selecting the Best Matching Candidate as Target}
\label{subsec:SelectTarget}

Given a selection of suitable candidate programs based on the correctness criterion, one of these programs has to be chosen as target solution, the one best suited for generating hints for the source program. This selection is done by calculating the distance between the source and all possible candidate programs, and then choosing the one with the smallest distance. Intuitively, the program with the smallest distance is most likely to represent a similar solution approach as used in the source program.

The distance between two programs is commonly calculated using their abstract
syntax trees. \scratch programs have some peculiarities over text-based
languages: While in text-based languages the order of all elements of a program
is given by the textual representation, the order of scripts and sprites in a
\scratch program is less clear: Scripts and sprites can be ordered and arranged
in arbitrary ways, and in the internal JSON representation of the program the
order of scripts is typically based on the order in which they were generated.
The pq-gram distance~\cite{augusten2005}, which is an efficient approximation
for the tree-edit distance, favors the existence of subtrees over the order of
these subtrees, and is therefore particularly suited for the application in the
\scratch context: It does not matter which sprite is added first or which
script is the first one in a sprite as long as it is there.

A pq-gram consists of an anchor node, $p-1$ of its ancestors, and
$q$ of its children (e.g., a 2-3 gram, has the direct parent of the
anchor, the anchor itself, and 3 of its children).
A pq-gram is represented as a node-tuple starting with the $p$ parent nodes,
with the anchor as last parent node, followed by the $q$ child nodes
$(a_1,...,a_{p},c_1,...,c_{q})$ (i.e., traversing the subtree in preorder). The
bag of all node-tuples of an AST $T$ is the pq-gram profile $P^{p,q}(T)$.

Given the pq-gram profiles for two trees $T_1$ and $T_2$, the pq-gram distance
is calculated as:
\begin{equation*}
 \Delta^{p,q}(T_1,T_2) = 1-2\frac{|P^{p,q}(T_1)\cap
P^{p,q}(T_2)|}{|P^{p,q}(T_1)|+|P^{p,q}(T_2)|} 
\end{equation*}
Thus, two trees with the same
pq-gram profile have distance 0 (note the pq-gram profiles are bags and not
sets); this, however, does not necessarily mean that the trees are equal but
only that they consist of the same subtrees. Higher $p$ values make the pq-gram
distance more sensitive to structural differences, whereas higher $q$ values
result in higher sensitivity for local changes in the leafs.

\catnip parses programs to their ASTs, calculates their pq-gram profiles, and
then calculates the pq-gram distances using these pq-gram profiles.
Finally, the \scratch project with the lowest pq-gram distance to the source is
selected as target. \catnip uses 2-3-grams by default, because having more children and thus block sequences in common seems most reasonable for the context~\cite{augusten2005,zimmerman2015}.

\subsection{Determining AST Differences}

Given a pair of pq-profiles for the source program and a chosen target program,
the next step is to identify the set $D_P$ of AST-nodes that exist only in the
source program (which represent candidates for deletions) and the set of
AST-nodes $A_P$ that exist only in the target program (which represent
candidates for additions).

This first requires matching the individual scripts of the two programs. Unlike
prior work~\cite{zimmerman2015}, \catnip exploits the program structure of
\scratch programs: Every AST for a \scratch project has the same basic
structure of a program that defines multiple actors (stage and any number of
sprites), each of which contains a set of scripts (a set of connected blocks,
which may be loose blocks, connected with a hat-block to denote an event, or
custom block definitions). In a first step, \catnip tries to match each actor
in the source project to an actor with the identical name in the target
program. If no exact match is found, then the pq-gram distance of the actor
subtrees is used to decide on the closest actor. If two actors have the same
distance one is selected randomly. Each actor can only be selected as target
once to prevent that two similar actors in the source are matched with the same target actor, thus resulting in hints that makes them functional clones.

Once actors have been matched, for each actor the pq-gram distance is used in order to determine, for each script in the source actor, the closest matching script in the target actor. 
If the source actor has more scripts than the target actor, then \catnip
iterates over the scripts of the target to find pairs; if the target actor has
more scripts than the source actor, then \catnip iterates over the scripts of
the source actor. During this iteration, each script is matched with the
closest script of the other actor, and once matched a script is not considered
for matching again. If there are unmatched scripts in the source actor, all
nodes of these scripts are candidates for deletion and added to $D_P$; if there
are unmatched scripts in the target actor, these are candidates for addition
and their event handler added to $A_P$. For each pair of matched scripts \catnip determines the intersection of AST-nodes; any nodes that occur only in the target are added as possible additions to $A_P$, and any nodes that exist only in the source are added to $D_P$ to denote possible deletions.

\subsection{Synthesizing Hints}
\label{subsec:SynthHints}

In order to produce understandable suggestions, a hint consists of the node to
be added or deleted as well as its context. Since scripts in \scratch can be
long, identifying where to add or delete a specific node may not be obvious.
While prior work~\cite{zimmerman2015} only considered the parent node in the
AST, this information may not be sufficient in the light of the often lengthy
scripts in \scratch. Therefore, a hint in \catnip consists of the node to be
added/deleted, its parent node, its $q-1$ sibling nodes (to both sides), as
well as the relevant actor and script. \catnip produces one deletion hint for
each node in $D_P$ and one addition hint for each node in $A_P$. The selection
of which of these hints to present to the user is deferred to the user
interface from which \catnip is invoked.

\section{Evaluation}\label{sec:evaluation}

To evaluate \catnip's ability to provide reasonable hints, we investigate the
following research questions:
\begin{compactdesc}
	\item \textbf{RQ1:} Can \catnip hints improve the correctness of student solutions?
	\item \textbf{RQ2:} How does the automated quality check of \catnip influence the hints?
\end{compactdesc}

\subsection{Experimental Setup}

\paragraph{Dataset:}
As a dataset for our evaluation we use the Fruit Catching game produced as part of two Scratch workshops with a sixth and a seventh grade class, respectively~\cite{stahlbauer2019testing}. The game consists of three sprites, a bowl that the player can move left or right with the cursor keys, and apples and bananas that drop from random positions at the top of the screen to the bottom. The aim of the game is to catch as much fruit as possible within a given time, without dropping any. There are 36 student implementations and 1 model solution. We furthermore use the \whisker test suite provided as part of the \whisker study~\cite{stahlbauer2019testing}, which contains 28 individual test cases checking different grading aspects of the solutions.

\paragraph{Procedure:}
We applied \catnip on each of the 36 student solutions in turn, using the remaining 35 student solutions and the model solution as a pool for hint generation. 
In actual use the student seeking help could specify which block they require a hint for.
In order to evaluate whether the hints generally point in the right direction without an actual user study, we consider \emph{all} hints produced by \catnip, and generate a version of the student program that applies all of them. Note that we only query \catnip once, i.e., we do not repeatedly request hints for code modified by hints. Since \catnip only gives recommendations at the level of blocks, specific values for numbers and strings were taken from the task specification.

To evaluate whether the hints have improved a solution, we run
\whisker~\cite{stahlbauer2019testing} on the student solution before and after
applying the hints, and compare the number of passed tests. We expect the
number of passing tests to increase if the hints improve the program. We
further use \litterbox~\cite{fraedrich2020} to extract metrics on the size and
quality of the programs before and after applying hints.
To answer RQ1, we set the threshold for programs acceptable as hint source to 90\%, such that only very good programs are used to generate hints (in this dataset, two programs satisfy this criterion). This resembles the use of hand-crafted or hand-selected solutions used in prior work.
To answer RQ2, we compare the quality of the hints produced by \catnip using other quality thresholds. In particular, we look at the results produced when using only the model solution, \emph{all} student solutions (i.e., a 0\% threshold) and 70\%, which will include more variations and results in 10 candidate programs.

\paragraph{Threats to Validity:} While our measurement of correctness tells us
whether \catnip provides hints that help move towards the right solution, the question whether these hints can help students who are stuck is an orthogonal question not answered by our experiments. We only used one task and student solutions from two classes as well as one test suite, so the results may not generalize to other classes or test suites. Nonetheless the used test suite was deemed suitable for automated grading in previous work~\cite{stahlbauer2019testing}.

\subsection{RQ1: Can \catnip hints improve the correctness of student solutions?}

\begin{figure}[t]
	\centering
	\includegraphics[width=1\columnwidth]{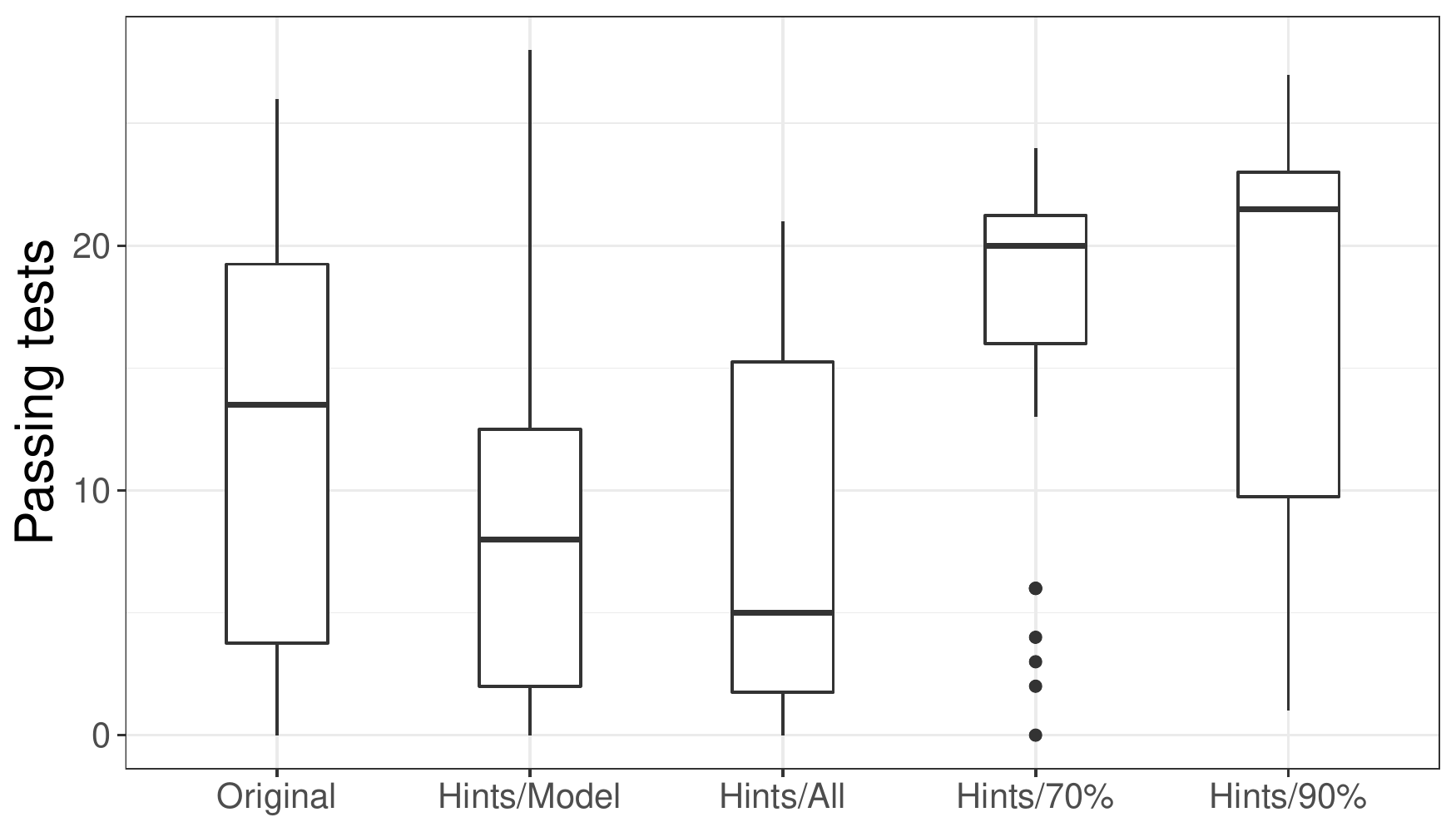}
	\vspace{-1em}
	\caption{\label{fig:passed_tests}Passed tests before and after hints are applied.}
\end{figure}

\begin{figure}[t]
	\centering
	\includegraphics[width=0.6\columnwidth]{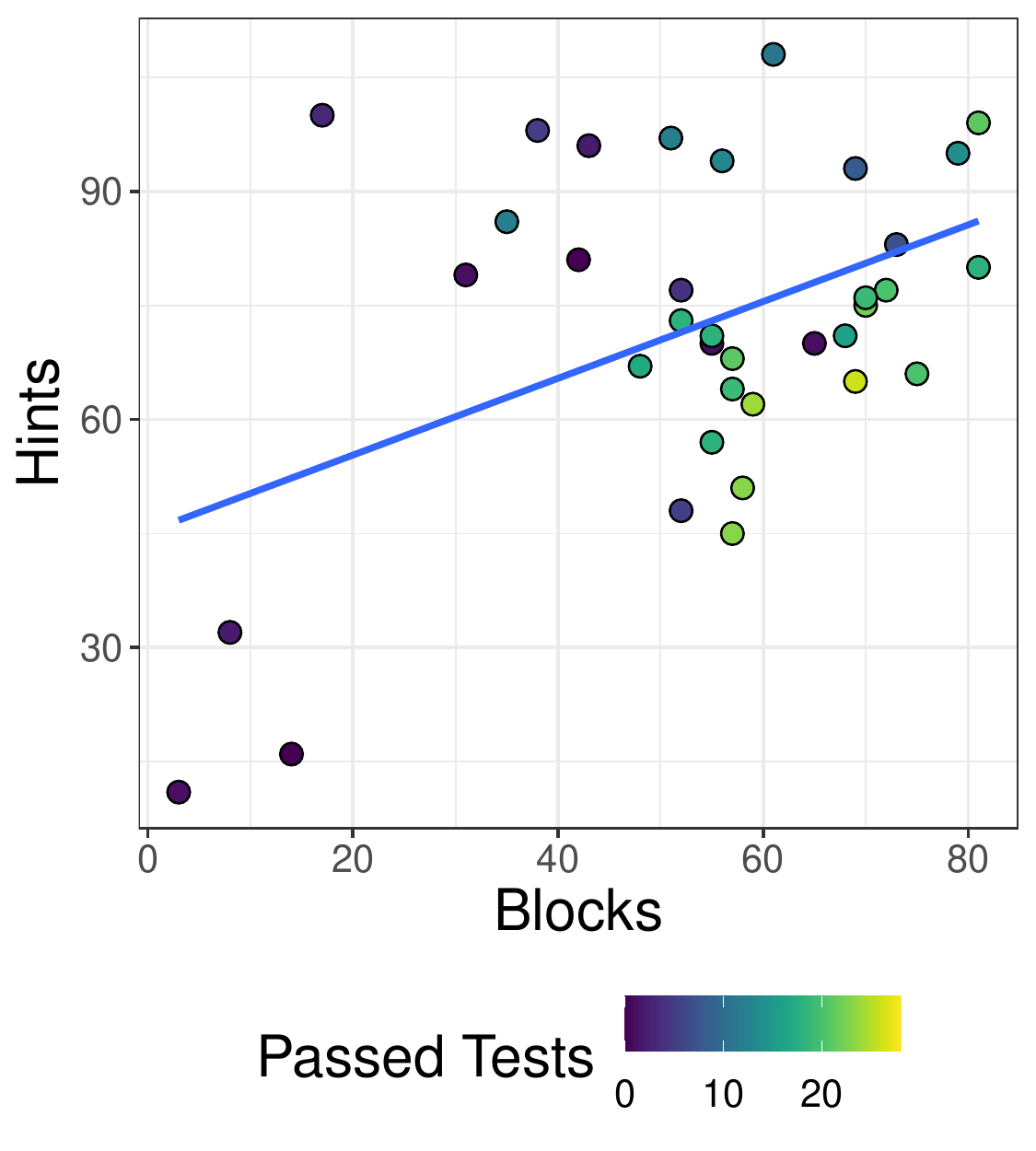}
	\vspace{-1em}
	\caption{\label{fig:blocks_vs_hints}Blocks vs hints generated.}
	\vspace{-1em}
\end{figure}

\begin{figure}[t]
	\centering
	\includegraphics[width=1\columnwidth]{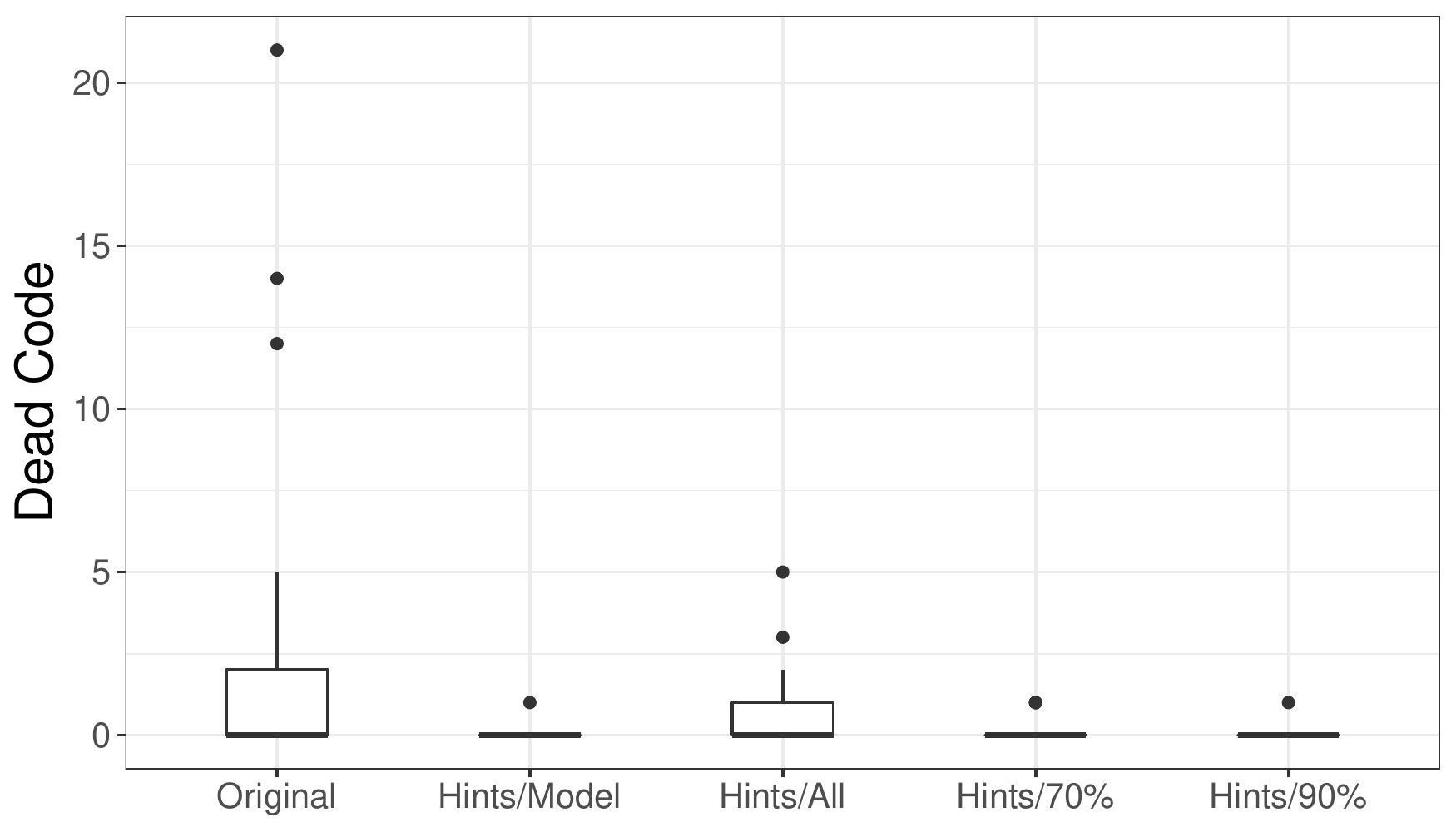}
	\caption{\label{fig:dead_code}Dead code before and after hints are applied.}
\end{figure}

\begin{figure}[t]
	\centering
	\includegraphics[width=1\columnwidth]{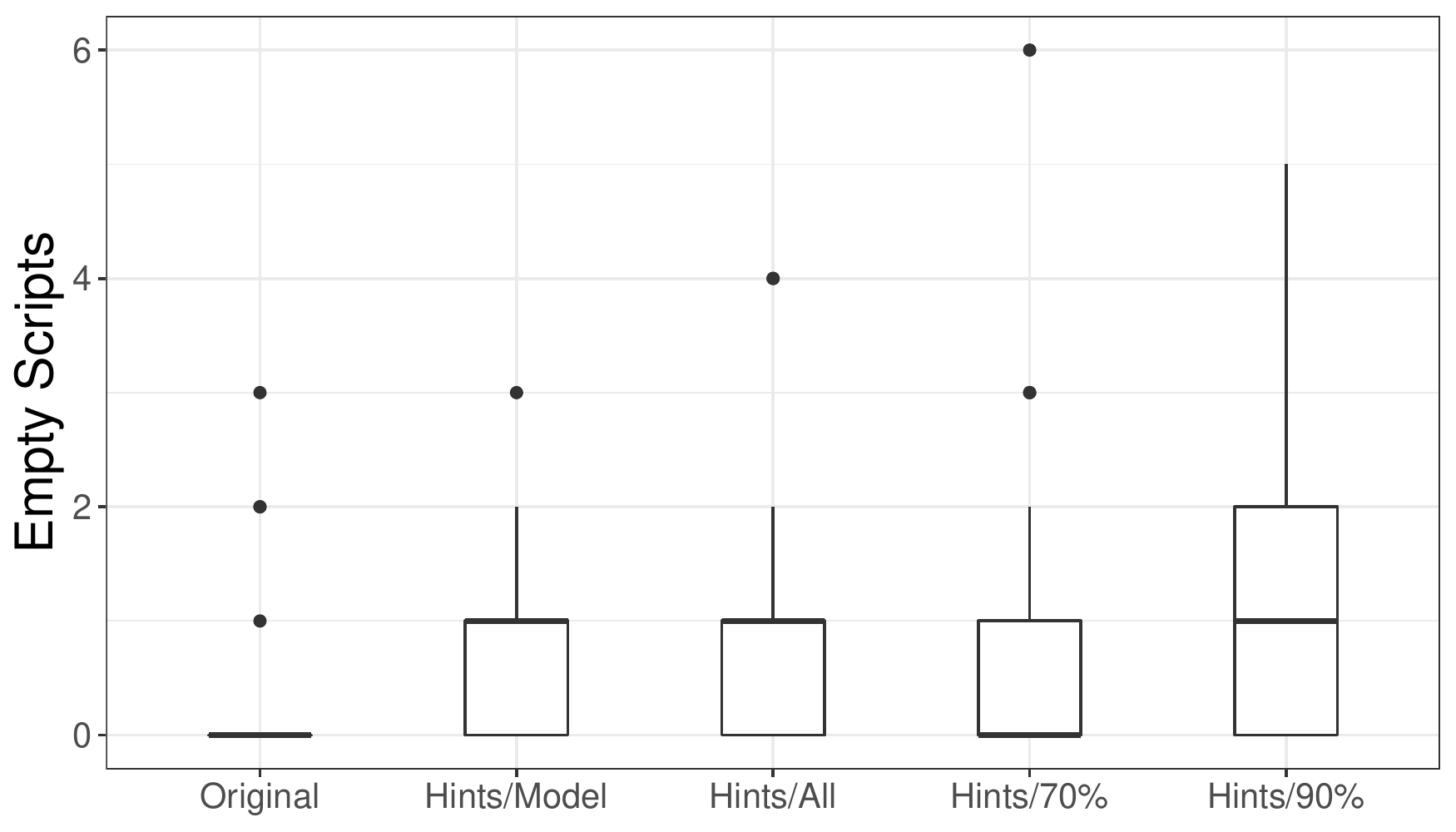}
	\caption{\label{fig:empty_script}Empty scripts before and after hints are applied.}
\end{figure}

\begin{figure}[t]
	\centering
	\includegraphics[width=1\columnwidth]{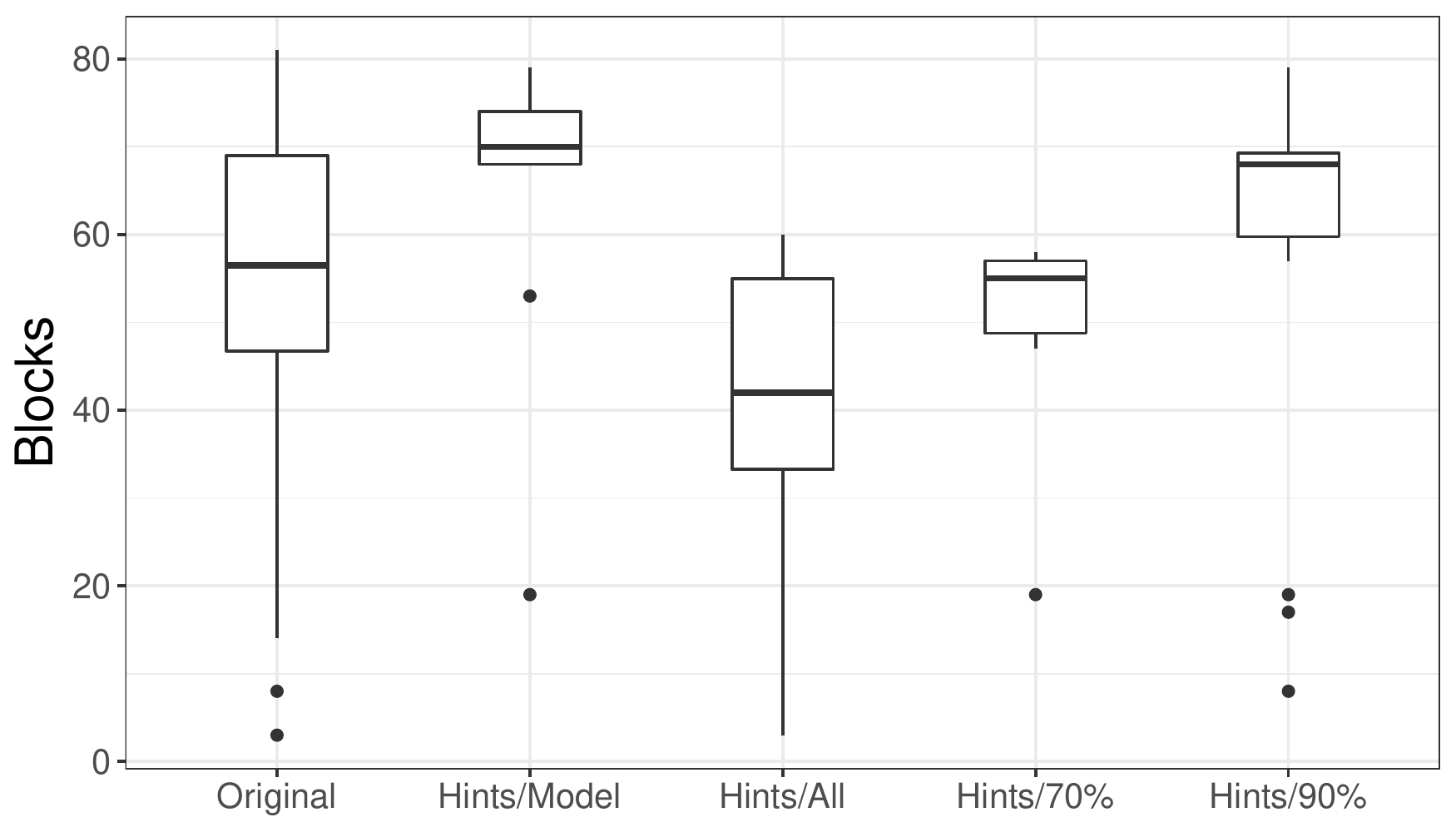}
	\caption{\label{fig:blocks}Blocks before and after hints are applied.}
\end{figure}

Figure \ref{fig:passed_tests} summarizes how many tests the solutions passed
before and after hints were applied. To answer RQ1, we only consider the
categories `Original' and `Hints/90\%', which represents \catnip's default
configuration. Integrating the hints produced by \catnip increases the number
of tests passed from an average of \meanPassedOriginal to \meanPassedNinety;
this difference is statistically significant according to a Mann-Whitney U test
($p = \passedOriginalNinety$, Vargha-Delaney effect size $\hat{A}_{12} =
\ApassedOriginalNinety$\footnote{$\hat{A}_{12}$ values less than 0.5 signify that the original program passes fewer tests, values greater than 0.5 signify the version with hints applied passes fewer tests.}). This clearly demonstrates that \catnip is able to
provide hints that lead to an improvement of the correctness of the student
solutions.

The number of passing tests represents a substantial increase in the
correctness of the student programs, and of course this cannot be achieved by a
single hint, but only by the combination of all generated hints as used in our
experiment. In general, the number of generated hints strongly depends on the
number of blocks that are already in the projects as shown in
Figure~\ref{fig:blocks_vs_hints} (Pearson's correlation coefficient
$\corrBlocksHints$, $p=\PcorrBlocksHints$), whereas there is no correlation
between the number of tests a program passes and the number of hints that are
generated for it (indicated with color shading in
Figure~\ref{fig:blocks_vs_hints}; correlation coefficient $\corrPassedHints$,
$p=\PcorrPassedHints$). In practice, a learner would not apply all hints, but
would, for example, request hints at specific locations. Note that invoking
\catnip iteratively more than once would likely further increase the number of
passing tests.

\begin{figure}[t]
	\begin{subfigure}[t]{0.4\columnwidth}
		\includegraphics[scale=0.37]{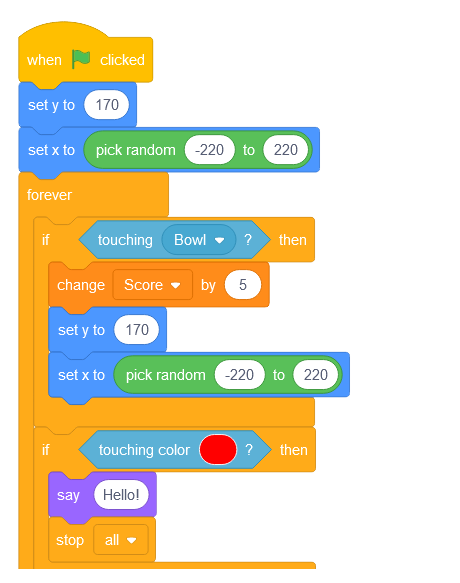}
		\caption{\label{fig:K7_S17_apple} Good student solution.}
	\end{subfigure}
	\hfill
	\begin{subfigure}[t]{0.58\columnwidth}
		\includegraphics[scale=0.37]{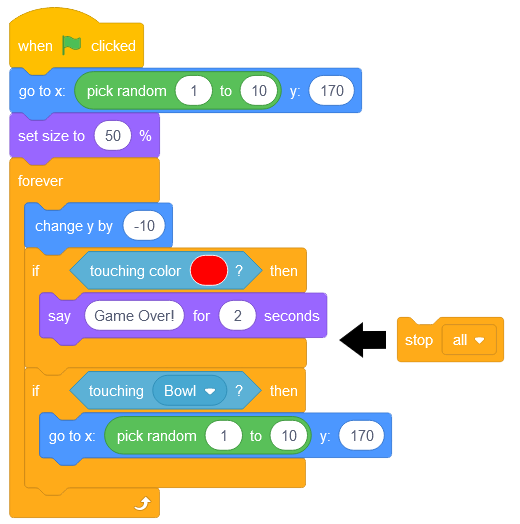}
		\caption{\label{fig:good_example_apple}Student solution receiving a hint.}
	\end{subfigure}
	\vspace{-1em}
	\caption{\label{fig:hint_from_student}The script for the apple is matched to the good solution, resulting in a hint to add a \textit{stop all} block.}
\end{figure}

As an example, Figure~\ref{fig:good_example_apple} shows a student
implementation of the apple sprite, which erroneously does not stop when
touching the red line. It is matched to the best student solution (shown in
Figure~\ref{fig:K7_S17_apple}), which itself is not perfect as `Hello' is said
instead of `Game over!', but still retains most of the necessary features for
the sprite. This match results in a hint suggesting to add a \textit{stop all}
block. Applying this hint to~\ref{fig:good_example_apple} improves the script
with the functionality to stop at the red line, thus making a further grading
test pass.

Figure~\ref{fig:dead_code} shows that the hints also lead to an improvement of
the general code quality, as the occurrence of ``dead code'' (i.e., partial
scripts or individual blocks lying around without being connected to
hat-blocks), is reduced on average from \meanDeadOriginal to \meanDeadNinety
($p~ \deadOriginalNinety$). On the other hand, Figure~\ref{fig:empty_script}
shows that \catnip often suggests to add new scripts, which initially contain
nothing but the event handler (increasing from an average of \meanEmptyOriginal
in to \meanEmptyNinety, $p~ \emptyOriginalNinety$). This is by design, as we
believe learners should not be provided with a model solution for an entire new
script, and \catnip's hints therefore just suggests what type of script should
be added. On balance, Figure~\ref{fig:blocks} suggests that hints tend to
suggest to add blocks.

\begin{figure}[t]
	\begin{subfigure}[t]{0.49\columnwidth}
		\includegraphics[scale=0.37]{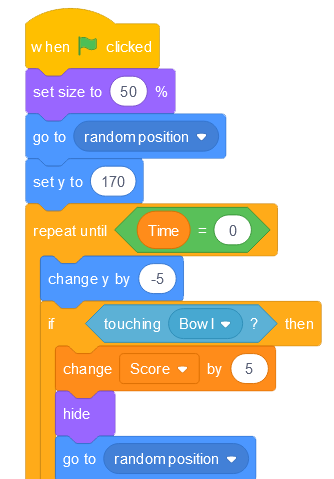}
		\caption{\label{fig:sample_apple}Model solution.}
	\end{subfigure}
	\hfill
	\begin{subfigure}[t]{0.49\columnwidth}
		\includegraphics[scale=0.37]{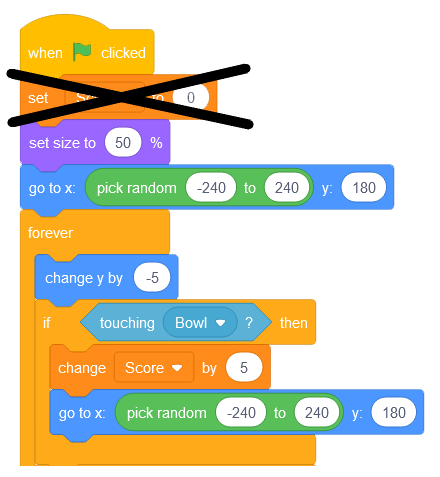}
		\caption{\label{fig:bad_example_apple}Student solution with hint.}
	\end{subfigure}
	\vspace{-1em}
	\caption{\label{fig:hint_from_sample}The script for the apple in the student solution is matched to the model solution. This results in a counterproductive hint to delete the \textit{set variable} block.}
\end{figure}

Although the hints improve most solutions, there are some exceptions where
hints negatively affect the functionality. This usually happens when the source
program is matched with a target program that uses a different solution
approach. For example, Figure~\ref{fig:bad_example_apple} shows an apple
implementation that was matched to the model solution shown in
Figure~\ref{fig:sample_apple}. The model solution handles the \textit{Score}
variable on the stage rather than the apple sprite, and \catnip suggests to
remove the block resetting the score. However, since the
student solution has no code on the stage, the score functionality is broken
after applying this hint.
This problem could have been avoided if the student solution had been matched to a target program using a more similar solution approach.
Interestingly, this problem particularly affects the only pupil solution
surpassing the 90\% criterion, as this only has the model solution to derive
hints from; the model solution, however, is structurally different, and the
hints attempt to remodel it closer to the model solution, after which only 12 tests (43\%) pass. 


\subsection{RQ2: How does the automated quality check of \catnip influence the hints?}

The results in RQ1 are based on using only solutions that pass 90\% of the tests as candidates for hint generation. In our dataset, only the model solution and one student solution matched this criterion. The aim of RQ2 is to investigate the influence of using student solutions and a quality criterion using automated tests.

When using only the model solution for hint generation, Figure~\ref{fig:passed_tests} shows that the hints (\emph{Hints/Model}) overall reduce the number of passing tests ($p = \passedOriginalModel$, $\hat{A}_{12}  = \ApassedOriginalModel$). That is, while there are hints that lead to an improvement, overall using just one model solution as candidate for hint generation is not recommendable, reinforcing the underlying idea of \catnip to use student solutions as well.
%


The larger increase in the number of blocks shown in Figure~\ref{fig:blocks}
suggests that the model solution uses a different approach than many students,
thus often resulting in the problem discussed above in
Figure~\ref{fig:hint_from_sample}. In particular Figure~\ref{fig:sample_apple}
shows the dependence of the apple movement and scoring from a correct
\textit{Time} reset, which many projects cannot fulfil after implementing the
hints derived from the model. This suggests that using only the model solution
is not ideal for individual feedback, as it encourages heavy remodelling and
guiding students to that very solution without considering progress made with a
different project structure.

However, using \emph{all} student solutions without any quality control is also
a bad idea as demonstrated in Figure~\ref{fig:passed_tests}, which shows an even lower
number of passing tests for \emph{Hints/All} ($p = \passedOriginalAll$,
$\hat{A}_{12} = \ApassedOriginalAll$). Very often, the pq-gram distance will
select other broken or incomplete programs as nearest solutions rather than an
actually helpful solution. This also leads to a decrease of the total number of blocks seen in Figure~\ref{fig:blocks}, which implies that most of the generated hints are deletions. The selection of programs that are not suited for hint generation also leads to dead code still being more present than in other configurations (see Figure~\ref{fig:dead_code}).

While it is clear that a quality threshold is necessary, it is less clear what
constitutes a good criterion. We therefore included a weaker 70\% correctness
criterion which leads to 10 solutions as candidate solutions for hint generation.
This does lead to a statistically significant increase in the number of passed
tests (Figure~\ref{fig:passed_tests}, configuration \emph{Hints/70\%}) ($p =
\passedOriginalSeventy$, $\hat{A}_{12} = \ApassedOriginalSeventy$) over the
original student solutions. However, compared to 90\% correctness there is no
significant difference ($p = \passedSeventyNinety$), with a slightly better
result for 90\% ($\hat{A}_{12} = \ApassedSeventyNinety$) as the 70\% criterion
includes incorrect solutions as candidate solutions for hint generation.

An interesting observation in this case is that the 70\% hints did not create
as many empty scripts as the 90\% hints (Figure~\ref{fig:empty_script}),
resulting in a significant difference in empty scripts between the 70\% and
90\% configurations ($p = \emptySeventyNinety$, $\hat{A}_{12} =
\AemptySeventyNinety$). Our conjecture is that more original projects were
matched with a target that fits their structure, and less new scripts had to be
created. In other words, the set of candidate solutions (= possible targets) is more diverse so
generated feedback can be tailored closer to an individual student project.
In general, the best correctness threshold will depend on the actual task, the programs in the pool of candidate solutions, the test suite, and the program for which hints are to be generated. 

\section{Conclusions}
\label{sec:conclusions}

Data-driven hint generation systems use candidate solutions for programming tasks
in order to generate suggestions how learners could continue solving their task. Creating or selecting suitable candidate solutions is challenging, but crucially important for generating truly helpful hints:
%
If a student attempts a valid solution path for which no model solution exists,
a data-driven hint generation system may provide counterproductive hints,
leading to unnecessary U-turns in students' learning paths. Students might
wrongly attribute this to missing capabilities, a situation that is unfair,
undesirable and may be harmful. On the other hand, if programs that use the
appropriate solution path, but are otherwise incorrect, are used for hint
generation, the resulting hints may be arbitrary or nonsensical
and thus lead to even more confusion.
%
In this paper we introduce \catnip, which is, to the best of our knowledge, the
first data-driven hint generation system for \scratch. \catnip uses automated
tests in order to select suitable solutions for hint generation, which makes it
easy to consider more candidate solutions, while ensuring that they are
sufficiently correct.

Our experiments with \catnip demonstrate how the use of student solutions
improves hints over the use of a single model solution, but also how the
quality of the student solutions is important in determining the validity of
the hints generated. Using \catnip with an appropriate correctness threshold on
an example dataset, the hints are of high quality and guide towards improved
functionality.
An important next step will be to integrate and evaluate the effects of using
\catnip in a real learning environment, to see whether students can use the
generated hints when they face problems. This also provides further
opportunities to improve \catnip, for example by combining deletion and
addition hints to replacement hints, by integrating Parsons problems~\cite{zhi2019} in hints, or by exploiting the automated tests
within the hint generation algorithm itself.
\catnip is available at: \url{https://github.com/se2p/catnip}.


\balance

\bibliographystyle{ACM-Reference-Format}
\bibliography{references}


\begin{thebibliography}{28}


\ifx \showCODEN    \undefined \def \showCODEN     #1{\unskip}     \fi
\ifx \showDOI      \undefined \def \showDOI       #1{#1}\fi
\ifx \showISBNx    \undefined \def \showISBNx     #1{\unskip}     \fi
\ifx \showISBNxiii \undefined \def \showISBNxiii  #1{\unskip}     \fi
\ifx \showISSN     \undefined \def \showISSN      #1{\unskip}     \fi
\ifx \showLCCN     \undefined \def \showLCCN      #1{\unskip}     \fi
\ifx \shownote     \undefined \def \shownote      #1{#1}          \fi
\ifx \showarticletitle \undefined \def \showarticletitle #1{#1}   \fi
\ifx \showURL      \undefined \def \showURL       {\relax}        \fi
\providecommand\bibfield[2]{#2}
\providecommand\bibinfo[2]{#2}
\providecommand\natexlab[1]{#1}
\providecommand\showeprint[2][]{arXiv:#2}

\bibitem[\protect\citeauthoryear{Augsten, B\"{o}hlen, and Gamper}{Augsten
  et~al\mbox{.}}{2005}]%
        {augusten2005}
\bibfield{author}{\bibinfo{person}{Nikolaus Augsten}, \bibinfo{person}{Michael
  B\"{o}hlen}, {and} \bibinfo{person}{Johann Gamper}.}
  \bibinfo{year}{2005}\natexlab{}.
\newblock \showarticletitle{Approximate Matching of Hierarchical Data Using
  pq-Grams}. In \bibinfo{booktitle}{\emph{Proceedings of the 31st International
  Conference on Very Large Data Bases}} (Trondheim, Norway)
  \emph{(\bibinfo{series}{VLDB '05})}. \bibinfo{publisher}{VLDB Endowment},
  \bibinfo{pages}{301–312}.
\newblock
\showISBNx{1595931546}


\bibitem[\protect\citeauthoryear{Bau, Gray, Kelleher, Sheldon, and Turbak}{Bau
  et~al\mbox{.}}{2017}]%
        {bau2017}
\bibfield{author}{\bibinfo{person}{David Bau}, \bibinfo{person}{Jeff Gray},
  \bibinfo{person}{Caitlin Kelleher}, \bibinfo{person}{Josh Sheldon}, {and}
  \bibinfo{person}{Franklyn Turbak}.} \bibinfo{year}{2017}\natexlab{}.
\newblock \showarticletitle{Learnable Programming: Blocks and Beyond}.
\newblock \bibinfo{journal}{\emph{Commun. ACM}} \bibinfo{volume}{60},
  \bibinfo{number}{6} (\bibinfo{date}{May} \bibinfo{year}{2017}),
  \bibinfo{pages}{72–80}.
\newblock
\showISSN{0001-0782}
\urldef\tempurl%
\url{https://doi.org/10.1145/3015455}
\showDOI{\tempurl}


\bibitem[\protect\citeauthoryear{Boe, Hill, Len, Dreschler, Conrad, and
  Franklin}{Boe et~al\mbox{.}}{2013}]%
        {boe2013}
\bibfield{author}{\bibinfo{person}{Bryce Boe}, \bibinfo{person}{Charlotte
  Hill}, \bibinfo{person}{Michelle Len}, \bibinfo{person}{Greg Dreschler},
  \bibinfo{person}{Phillip Conrad}, {and} \bibinfo{person}{Diana Franklin}.}
  \bibinfo{year}{2013}\natexlab{}.
\newblock \showarticletitle{Hairball: Lint-inspired static analysis of scratch
  projects}.
\newblock \bibinfo{journal}{\emph{SIGCSE 2013 - Proceedings of the 44th ACM
  Technical Symposium on Computer Science Education}},
  \bibinfo{pages}{215--220}.
\newblock
\urldef\tempurl%
\url{https://doi.org/10.1145/2445196.2445265}
\showDOI{\tempurl}


\bibitem[\protect\citeauthoryear{Collins and Kapur}{Collins and Kapur}{2014}]%
        {collins_kapur_2014}
\bibfield{author}{\bibinfo{person}{Allan Collins} {and} \bibinfo{person}{Manu
  Kapur}.} \bibinfo{year}{2014}\natexlab{}.
\newblock \bibinfo{booktitle}{\emph{Cognitive Apprenticeship}
  (\bibinfo{edition}{2} ed.)}.
\newblock \bibinfo{publisher}{Cambridge University Press},
  \bibinfo{pages}{109–127}.
\newblock
\urldef\tempurl%
\url{https://doi.org/10.1017/CBO9781139519526.008}
\showDOI{\tempurl}


\bibitem[\protect\citeauthoryear{Fowler}{Fowler}{1999}]%
        {fowler1999}
\bibfield{author}{\bibinfo{person}{Martin Fowler}.}
  \bibinfo{year}{1999}\natexlab{}.
\newblock \bibinfo{booktitle}{\emph{Refactoring: Improving the Design of
  Existing Code}}.
\newblock \bibinfo{publisher}{Addison-Wesley}, \bibinfo{address}{Boston, MA,
  USA}.
\newblock
\showISBNx{0-201-48567-2}


\bibitem[\protect\citeauthoryear{Frädrich, Obermüller, Körber, Heuer, and
  Fraser}{Frädrich et~al\mbox{.}}{2020}]%
        {fraedrich2020}
\bibfield{author}{\bibinfo{person}{Christoph Frädrich},
  \bibinfo{person}{Florian Obermüller}, \bibinfo{person}{Nina Körber},
  \bibinfo{person}{Ute Heuer}, {and} \bibinfo{person}{Gordon Fraser}.}
  \bibinfo{year}{2020}\natexlab{}.
\newblock \showarticletitle{Common Bugs in Scratch Programs}. In
  \bibinfo{booktitle}{\emph{Proceedings of the 2020 ACM Conference on
  Innovation and Technology in Computer Science Education}} (Trondheim, Norway)
  \emph{(\bibinfo{series}{ITiCSE '20})}. \bibinfo{pages}{89--95}.
\newblock
\urldef\tempurl%
\url{https://doi.org/10.1145/3341525.3387389}
\showDOI{\tempurl}


\bibitem[\protect\citeauthoryear{Garcia, Harvey, and Barnes}{Garcia
  et~al\mbox{.}}{2015}]%
        {garcia2015}
\bibfield{author}{\bibinfo{person}{Dan Garcia}, \bibinfo{person}{Brian Harvey},
  {and} \bibinfo{person}{Tiffany Barnes}.} \bibinfo{year}{2015}\natexlab{}.
\newblock \showarticletitle{The Beauty and Joy of Computing}.
\newblock \bibinfo{journal}{\emph{ACM Inroads}} \bibinfo{volume}{6},
  \bibinfo{number}{4} (\bibinfo{date}{Nov.} \bibinfo{year}{2015}),
  \bibinfo{pages}{71–79}.
\newblock
\showISSN{2153-2184}
\urldef\tempurl%
\url{https://doi.org/10.1145/2835184}
\showDOI{\tempurl}


\bibitem[\protect\citeauthoryear{Hansen and Eddy}{Hansen and Eddy}{2007}]%
        {hansen2007}
\bibfield{author}{\bibinfo{person}{Stuart Hansen} {and} \bibinfo{person}{Erica
  Eddy}.} \bibinfo{year}{2007}\natexlab{}.
\newblock \showarticletitle{Engagement and Frustration in Programming
  Projects}. In \bibinfo{booktitle}{\emph{Proceedings of the 38th SIGCSE
  Technical Symposium on Computer Science Education}} (Covington, Kentucky,
  USA) \emph{(\bibinfo{series}{SIGCSE '07})}. \bibinfo{publisher}{Association
  for Computing Machinery}, \bibinfo{address}{New York, NY, USA},
  \bibinfo{pages}{271–275}.
\newblock
\showISBNx{1595933611}
\urldef\tempurl%
\url{https://doi.org/10.1145/1227310.1227407}
\showDOI{\tempurl}


\bibitem[\protect\citeauthoryear{Hermans and Aivaloglou}{Hermans and
  Aivaloglou}{2016}]%
        {hermans2016b}
\bibfield{author}{\bibinfo{person}{Felienne Hermans} {and}
  \bibinfo{person}{Efthimia Aivaloglou}.} \bibinfo{year}{2016}\natexlab{}.
\newblock \showarticletitle{Do code smells hamper novice programming? A
  controlled experiment on Scratch programs}. In \bibinfo{booktitle}{\emph{2016
  IEEE 24th International Conference on Program Comprehension (ICPC)}}.
  \bibinfo{pages}{1--10}.
\newblock
\showISSN{null}
\urldef\tempurl%
\url{https://doi.org/10.1109/ICPC.2016.7503706}
\showDOI{\tempurl}


\bibitem[\protect\citeauthoryear{Hermans, Stolee, and Hoepelman}{Hermans
  et~al\mbox{.}}{2016}]%
        {hermans2016a}
\bibfield{author}{\bibinfo{person}{Felienne Hermans},
  \bibinfo{person}{Kathryn~T. Stolee}, {and} \bibinfo{person}{David
  Hoepelman}.} \bibinfo{year}{2016}\natexlab{}.
\newblock \showarticletitle{Smells in Block-Based Programming Languages}. In
  \bibinfo{booktitle}{\emph{2016 {{IEEE Symposium}} on {{Visual Languages}} and
  {{Human}}-{{Centric Computing}} ({{VL}}/{{HCC}})}} ({Cambridge, United
  Kingdom}, 2016-09). \bibinfo{publisher}{{IEEE}}, \bibinfo{pages}{68--72}.
\newblock
\showISBNx{978-1-5090-0252-8}
\urldef\tempurl%
\url{https://doi.org/10.1109/VLHCC.2016.7739666}
\showDOI{\tempurl}


\bibitem[\protect\citeauthoryear{Hovemeyer and Pugh}{Hovemeyer and
  Pugh}{2004}]%
        {hovemeyer2004}
\bibfield{author}{\bibinfo{person}{David Hovemeyer} {and}
  \bibinfo{person}{William Pugh}.} \bibinfo{year}{2004}\natexlab{}.
\newblock \showarticletitle{Finding Bugs is Easy}.
\newblock \bibinfo{journal}{\emph{SIGPLAN Not.}} \bibinfo{volume}{39},
  \bibinfo{number}{12} (\bibinfo{date}{Dec.} \bibinfo{year}{2004}),
  \bibinfo{pages}{92–106}.
\newblock
\showISSN{0362-1340}
\urldef\tempurl%
\url{https://doi.org/10.1145/1052883.1052895}
\showDOI{\tempurl}


\bibitem[\protect\citeauthoryear{Johnson}{Johnson}{2016}]%
        {johnson2016itch}
\bibfield{author}{\bibinfo{person}{David~E Johnson}.}
  \bibinfo{year}{2016}\natexlab{}.
\newblock \showarticletitle{ITCH: Individual Testing of Computer Homework for
  Scratch Assignments}. In \bibinfo{booktitle}{\emph{Proceedings of the 47th
  ACM Technical Symposium on Computing Science Education}}. ACM,
  \bibinfo{pages}{223--227}.
\newblock


\bibitem[\protect\citeauthoryear{Maloney, Resnick, Rusk, Silverman, and
  Eastmond}{Maloney et~al\mbox{.}}{2010}]%
        {maloney2010}
\bibfield{author}{\bibinfo{person}{John Maloney}, \bibinfo{person}{Mitchel
  Resnick}, \bibinfo{person}{Natalie Rusk}, \bibinfo{person}{Brian Silverman},
  {and} \bibinfo{person}{Evelyn Eastmond}.} \bibinfo{year}{2010}\natexlab{}.
\newblock \showarticletitle{The Scratch Programming Language and Environment}.
\newblock \bibinfo{journal}{\emph{ACM Transactions on Computing Education
  (TOCE)}}  \bibinfo{volume}{10} (\bibinfo{date}{11} \bibinfo{year}{2010}),
  \bibinfo{pages}{16}.
\newblock
\urldef\tempurl%
\url{https://doi.org/10.1145/1868358.1868363}
\showDOI{\tempurl}


\bibitem[\protect\citeauthoryear{McGill and Decker}{McGill and Decker}{2020}]%
        {mcgill2020}
\bibfield{author}{\bibinfo{person}{Monica~M. McGill} {and}
  \bibinfo{person}{Adrienne Decker}.} \bibinfo{year}{2020}\natexlab{}.
\newblock \showarticletitle{Tools, Languages, and Environments Used in Primary
  and Secondary Computing Education}. In \bibinfo{booktitle}{\emph{Proceedings
  of the 2020 ACM Conference on Innovation and Technology in Computer Science
  Education}} (Trondheim, Norway) \emph{(\bibinfo{series}{ITiCSE '20})}.
  \bibinfo{publisher}{Association for Computing Machinery},
  \bibinfo{address}{New York, NY, USA}, \bibinfo{pages}{103–109}.
\newblock
\showISBNx{9781450368742}
\urldef\tempurl%
\url{https://doi.org/10.1145/3341525.3387365}
\showDOI{\tempurl}


\bibitem[\protect\citeauthoryear{Moreno-Le{\'o}n and Robles}{Moreno-Le{\'o}n
  and Robles}{2014}]%
        {moreno2014}
\bibfield{author}{\bibinfo{person}{Jes{\'u}s Moreno-Le{\'o}n} {and}
  \bibinfo{person}{Gregorio Robles}.} \bibinfo{year}{2014}\natexlab{}.
\newblock \showarticletitle{Automatic detection of bad programming habits in
  scratch: A preliminary study}. In \bibinfo{booktitle}{\emph{2014 IEEE
  Frontiers in Education Conference (FIE) Proceedings}}. \bibinfo{pages}{1--4}.
\newblock
\showISSN{2377-634X}
\urldef\tempurl%
\url{https://doi.org/10.1109/FIE.2014.7044055}
\showDOI{\tempurl}


\bibitem[\protect\citeauthoryear{Price, Zhi, and Barnes}{Price
  et~al\mbox{.}}{2017b}]%
        {price2017b}
\bibfield{author}{\bibinfo{person}{Thomas Price}, \bibinfo{person}{Rui Zhi},
  {and} \bibinfo{person}{Tiffany Barnes}.} \bibinfo{year}{2017}\natexlab{b}.
\newblock \showarticletitle{Evaluation of a Data-driven Feedback Algorithm for
  Open-ended Programming}.
\newblock


\bibitem[\protect\citeauthoryear{Price, Dong, and Barnes}{Price
  et~al\mbox{.}}{2016}]%
        {price2016}
\bibfield{author}{\bibinfo{person}{Thomas~W. Price}, \bibinfo{person}{Yihuan
  Dong}, {and} \bibinfo{person}{Tiffany Barnes}.}
  \bibinfo{year}{2016}\natexlab{}.
\newblock \showarticletitle{Generating Data-driven Hints for Open-ended
  Programming}. In \bibinfo{booktitle}{\emph{EDM}}.
\newblock


\bibitem[\protect\citeauthoryear{Price, Dong, and Lipovac}{Price
  et~al\mbox{.}}{2017a}]%
        {price2017a}
\bibfield{author}{\bibinfo{person}{Thomas~W. Price}, \bibinfo{person}{Yihuan
  Dong}, {and} \bibinfo{person}{Dragan Lipovac}.}
  \bibinfo{year}{2017}\natexlab{a}.
\newblock \showarticletitle{ISnap: Towards Intelligent Tutoring in Novice
  Programming Environments}. In \bibinfo{booktitle}{\emph{Proceedings of the
  2017 ACM SIGCSE Technical Symposium on Computer Science Education}} (Seattle,
  Washington, USA) \emph{(\bibinfo{series}{SIGCSE '17})}.
  \bibinfo{publisher}{Association for Computing Machinery},
  \bibinfo{address}{New York, NY, USA}, \bibinfo{pages}{483–488}.
\newblock
\showISBNx{9781450346986}
\urldef\tempurl%
\url{https://doi.org/10.1145/3017680.3017762}
\showDOI{\tempurl}


\bibitem[\protect\citeauthoryear{Reiser and Tabak}{Reiser and Tabak}{2014}]%
        {reiser_tabak_2014}
\bibfield{author}{\bibinfo{person}{Brian~J. Reiser} {and} \bibinfo{person}{Iris
  Tabak}.} \bibinfo{year}{2014}\natexlab{}.
\newblock \bibinfo{booktitle}{\emph{Scaffolding} (\bibinfo{edition}{2} ed.)}.
\newblock \bibinfo{publisher}{Cambridge University Press},
  \bibinfo{pages}{44–62}.
\newblock
\urldef\tempurl%
\url{https://doi.org/10.1017/CBO9781139519526.005}
\showDOI{\tempurl}


\bibitem[\protect\citeauthoryear{Renninger}{Renninger}{2010}]%
        {renninger2010working}
\bibfield{author}{\bibinfo{person}{K~Ann Renninger}.}
  \bibinfo{year}{2010}\natexlab{}.
\newblock \showarticletitle{Working with and cultivating the development of
  interest, self-efficacy, and self-regulation.}
\newblock \bibinfo{journal}{\emph{Innovations in Educational Psychology}}
  (\bibinfo{year}{2010}), \bibinfo{pages}{107}.
\newblock


\bibitem[\protect\citeauthoryear{Robins}{Robins}{2019}]%
        {robins_2019}
\bibfield{author}{\bibinfo{person}{Anthony~V. Robins}.}
  \bibinfo{year}{2019}\natexlab{}.
\newblock \bibinfo{booktitle}{\emph{Novice Programmers and Introductory
  Programming}}.
\newblock \bibinfo{publisher}{Cambridge University Press},
  \bibinfo{pages}{327–376}.
\newblock
\urldef\tempurl%
\url{https://doi.org/10.1017/9781108654555.013}
\showDOI{\tempurl}


\bibitem[\protect\citeauthoryear{Singh, Gulwani, and Solar-Lezama}{Singh
  et~al\mbox{.}}{2012}]%
        {singh2012}
\bibfield{author}{\bibinfo{person}{Rishabh Singh}, \bibinfo{person}{Sumit
  Gulwani}, {and} \bibinfo{person}{Armando Solar-Lezama}.}
  \bibinfo{year}{2012}\natexlab{}.
\newblock \showarticletitle{Automated Feedback Generation for Introductory
  Programming Assignments}.
\newblock \bibinfo{journal}{\emph{ACM SIGPLAN Notices}}  \bibinfo{volume}{48}
  (\bibinfo{date}{04} \bibinfo{year}{2012}).
\newblock
\urldef\tempurl%
\url{https://doi.org/10.1145/2491956.2462195}
\showDOI{\tempurl}


\bibitem[\protect\citeauthoryear{Stahlbauer, Frädrich, and Fraser}{Stahlbauer
  et~al\mbox{.}}{2020}]%
        {stahlbauer2020}
\bibfield{author}{\bibinfo{person}{Andreas Stahlbauer},
  \bibinfo{person}{Christoph Frädrich}, {and} \bibinfo{person}{Gordon
  Fraser}.} \bibinfo{year}{2020}\natexlab{}.
\newblock \showarticletitle{{Verified from Scratch: Program Analysis for
  Learners’ Programs}}. In \bibinfo{booktitle}{\emph{In Proceedings of the
  International Conference on Automated Software Engineering (ASE)}}.
  \bibinfo{publisher}{{IEEE}}.
\newblock


\bibitem[\protect\citeauthoryear{Stahlbauer, Kreis, and Fraser}{Stahlbauer
  et~al\mbox{.}}{2019}]%
        {stahlbauer2019testing}
\bibfield{author}{\bibinfo{person}{Andreas Stahlbauer}, \bibinfo{person}{Marvin
  Kreis}, {and} \bibinfo{person}{Gordon Fraser}.}
  \bibinfo{year}{2019}\natexlab{}.
\newblock \showarticletitle{Testing scratch programs automatically}. In
  \bibinfo{booktitle}{\emph{Proceedings of the 2019 27th ACM Joint Meeting on
  European Software Engineering Conference and Symposium on the Foundations of
  Software Engineering}}. \bibinfo{pages}{165--175}.
\newblock


\bibitem[\protect\citeauthoryear{Techapalokul and Tilevich}{Techapalokul and
  Tilevich}{2017a}]%
        {techapaloku2017b}
\bibfield{author}{\bibinfo{person}{Peeratham Techapalokul} {and}
  \bibinfo{person}{Eli Tilevich}.} \bibinfo{year}{2017}\natexlab{a}.
\newblock \showarticletitle{Quality Hound — An online code smell analyzer for
  scratch programs}. In \bibinfo{booktitle}{\emph{2017 IEEE Symposium on Visual
  Languages and Human-Centric Computing (VL/HCC)}}. \bibinfo{pages}{337--338}.
\newblock
\showISSN{1943-6106}
\urldef\tempurl%
\url{https://doi.org/10.1109/VLHCC.2017.8103498}
\showDOI{\tempurl}


\bibitem[\protect\citeauthoryear{Techapalokul and Tilevich}{Techapalokul and
  Tilevich}{2017b}]%
        {techapalokul2017a}
\bibfield{author}{\bibinfo{person}{Peeratham Techapalokul} {and}
  \bibinfo{person}{Eli Tilevich}.} \bibinfo{year}{2017}\natexlab{b}.
\newblock \showarticletitle{Understanding Recurring Quality Problems and Their
  Impact on Code Sharing in Block-Based Software}. In
  \bibinfo{booktitle}{\emph{2017 {{IEEE Symposium}} on {{Visual Languages}} and
  {{Human}}-{{Centric Computing}} ({{VL}}/{{HCC}})}} ({Raleigh, NC, USA},
  2017-10). \bibinfo{publisher}{{IEEE}}, \bibinfo{pages}{43--51}.
\newblock
\showISBNx{978-1-5386-0443-4}
\urldef\tempurl%
\url{https://doi.org/10.1109/VLHCC.2017.8103449}
\showDOI{\tempurl}


\bibitem[\protect\citeauthoryear{Zhi, Chi, Barnes, and Price}{Zhi
  et~al\mbox{.}}{2019}]%
        {zhi2019}
\bibfield{author}{\bibinfo{person}{Rui Zhi}, \bibinfo{person}{Min Chi},
  \bibinfo{person}{Tiffany Barnes}, {and} \bibinfo{person}{Thomas~W. Price}.}
  \bibinfo{year}{2019}\natexlab{}.
\newblock \showarticletitle{Evaluating the Effectiveness of Parsons Problems
  for Block-Based Programming}. In \bibinfo{booktitle}{\emph{Proceedings of the
  2019 ACM Conference on International Computing Education Research}} (Toronto
  ON, Canada) \emph{(\bibinfo{series}{ICER '19})}.
  \bibinfo{publisher}{Association for Computing Machinery},
  \bibinfo{address}{New York, NY, USA}, \bibinfo{pages}{51–59}.
\newblock
\showISBNx{9781450361859}
\urldef\tempurl%
\url{https://doi.org/10.1145/3291279.3339419}
\showDOI{\tempurl}


\bibitem[\protect\citeauthoryear{{Zimmerman} and {Rupakheti}}{{Zimmerman} and
  {Rupakheti}}{2015}]%
        {zimmerman2015}
\bibfield{author}{\bibinfo{person}{K. {Zimmerman}} {and} \bibinfo{person}{C.~R.
  {Rupakheti}}.} \bibinfo{year}{2015}\natexlab{}.
\newblock \showarticletitle{An Automated Framework for Recommending Program
  Elements to Novices (N)}. In \bibinfo{booktitle}{\emph{2015 30th IEEE/ACM
  International Conference on Automated Software Engineering (ASE)}}.
  \bibinfo{pages}{283--288}.
\newblock


\end{thebibliography}


\end{document}